# Machine Learning vs. Deep Learning in 5G Networks – A Comparison of Scientific Impact

**Ilker TURKER [1*], Serhat Orkun TAN [2]**

[1] Karabuk University, Faculty of Engineering, Department of Computer Engineering (iturker@karabuk.edu.tr)
[2] Karabuk University, Faculty of Technology, Department of Electrical Engineering

**ABSTRACT**

Introduction of fifth generation (5G) wireless network technology has matched the crucial need for high capacity and speed needs of the new generation mobile applications. Recent advances in Artificial Intelligence (AI) also empowered 5G cellular networks with two mainstreams as machine learning (ML) and deep learning (DL) techniques.

Our study aims to uncover the differences in scientific impact for these two techniques by the means of statistical bibliometrics. The performed analysis includes citation performance with respect to indexing types, funding availability, journal or conference publishing options together with distributions of these metrics along years to evaluate the popularity trends in a detailed manner.

Web of Science (WoS) database host 2245 papers for ML and 1407 papers for DL-related studies. DL studies, starting with 9% rate in 2013, has reached to 45% rate in 2022 among all DL and ML-related studies. Results related to scientific impact indicate that DL studies get slightly more average normalized citation (2.256) compared to ML studies (2.118) in 5G, while SCI-Expanded indexed papers in both sides tend to have similar citation performance (3.165 and 3.162 respectively). ML-related studies those are indexed in ESCI show twice citation performance compared to DL. Conference papers in DL domain and journal papers in ML domain are superior in scientific interest to their counterparts with minor differences. Highest citation performance for ML studies is achieved for year 2014, while this peak is observed for 2017 for DL studies. We can conclude that both publication and citation rate for DL-related papers tend to increase and outperform ML-based studies in 5G domain by the means of citation metrics.

## 1. INTRODUCTION

Machine learning (ML) describes tools for imitating human perception by the computers, employing a variety of algorithms. It is combined with powerful data processing, with a learning procedure provided by a set of training data from the field. Many techniques from various domains such as statistics, pattern recognition, database, information retrieval, algorithms, high performance computing etc. are incorporated to form this subfield [1-3]. Tools introduced by ML environment enable scientists to make classification, clustering and associate analysis in many fields. ML is proved to present robust and efficient algorithms assisting in making or supporting decisions after a big data driven training phase [4]. These studies include many fields such as medicine [5, 6], biomedical science [7, 8], telecommunications [9], cyber-security [10], finance [11], mechanics [12], machining [13], material design [14] etc.

Over the course of the last decade, the community witnessed a sharp development in Deep Learning (DL) and Deep Neural Networks (DNN), a deep architecture of Artificial Neural Networks (ANN) having capabilities of extracting features and learning from large-scale data sources with high precision [15]. Deep Belief Networks (DBN), Restricted Boltzman Machines (RBM), Convolutional Neural Networks (CNN), Recurrent Neural Networks (RNN), Gated Recurrent Units (GRU), Long-Short Term Memory Networks (LSTM), Autoencoders (AE), Generative Adversarial Networks (GAN) are a variety of known DL models frequently used in current studies [16]. DL techniques emerge to provide significantly higher performance compared to traditional methods [17], within the scope covered by ML-driven studies mentioned above.

The flexible structure of DL architectures enables them to be modified to suit a wide variety of problems across all three types of ML techniques. DNNs typically consist of more than one hidden layer, designed to consist of deeply nested network layers [18]. Furthermore, they are composed of advanced neurons compared to simple ANNs. This superior structure leads to making advanced operations like convolutions or multiple activations in a single neuron rather than employing simple activation function as in ANNs. Taking advantage of this superiority, DNNs can process raw input data and discover the interesting patterns automatically to perform the desired learning task [19].

The main difference between ML and DL lies in the deep architecture of DL giving the capability of discovering features from raw data, while ML iteratively learns from training data after a feature selection procedure in an automated or handcrafted manner. ML algorithms have been successfully applied to many fields especially for limited data occasions, while deep learning models outperform the shallow ML algorithms especially in domains with relatively high dimensional data such as in text, image, video and audio format [20]. On the other hand, in case of limited data availability and relatively low dimensions, ML algorithms can still outperform deep learning models [21].

Fifth generation (5G) mobile networks aim at matching the high data transfer demand that contemporarily drives the improvements in technology and modern life. A key approach followed by network operators is adopting cloud-computing techniques and beyond intelligent technologies to improve its data transfer capacity and QoS. Advances in adopting artificial intelligence (AI) methods in 5G successfully provide performance improvements since 5G evolves to low-latency, high-availability, and high-bandwidth communication, thanks to delay-sensitive applications. Machine learning (ML) and deep learning (DL) represent the two main streams of AI in 5G networks (Huang, Guo et al. 2019, McClellan, Cervelló-Pastor et al. 2020, Santos, Endo et al. 2020).

The current study aims to provide a bibliometric view on the literature of 5G networks scope in a comparative manner. This aim will be achieved with a data-driven approach by differentiating the articles whether they use ML or DL-based methods. We use bibliometric data provided by the Web of Science (WoS) including more than 60 attributes such as language, document type, citation and usage count, publication year, page count, indexing information, research area etc. This variety of attributes enables a broad statistical comparison on which approach has the greater impact within the 5G networks scope.

## 2. DATA AND METHODS

The dataset is provided from the WoS Core Collection, utilizing the user interface which enables writing queries and listing the related studies. The interface also enables batch downloading the results in bins of 1000 records. The provided results include many attributes of the publications including publication type, years, research field, authors, affiliations, journal or conference titles, publishers, WoS index etc. We performed queries for the two counterparts of the comparative study as:

*Machine Learning :* ("5G") AND ("machine learning" OR "data mining" OR "clustering" OR "decision trees" OR "support vector machines" OR "SVM" OR "random forest" OR "k-nearest neighbors" OR "k-NN" OR "kNN" OR "naive Bayes" OR "association rule mining")

*Deep Learning:* ("5G") AND ("deep learning" OR "neural networks" OR "convolutional neural network" OR "CNN" OR "Recurrent Neural Networks" OR "RNN" OR "Deep Neural Networks" OR "DNN" OR "Long-Short Term Memory" OR "LSTM").

Results of the mentioned searches include all the studies indexed under WoS database for all available years and WoS categories. Once retrieving the data, we first spotted a general outline of some specific fields as publication years, document types, affiliations, publication titles, publishers, open access, countries regions, language, WoS Index in a comparative graphical presentation for the two main search results. These graphics include aggregate statistics for the mentioned attributes mostly for the selected attributes with nominal values. Second mainstream of the outputs provided are based on numeric attributes of the results, those are mainly related with the performance metrics of the studies such as citation count, usage count and page count. Analysis related with this section involve in statistical inferences from probability distributions.

## 3. RESULTS AND DISCUSSION

The analysis of data in a comparative manner is presented in this section in two subsections. The first subsection provides a basic statistical view on the available attributes, while the second subsection focuses on scientific impact, based on citation and usage statistics.

### 3.1. Basic Statistical Results

This subsection gives a comparison for DL and ML studies conducted for 5G Networks in yearly publication count, document types, publication titles (sources), funding availability, research areas, countries and indexing information basis.

#### 3.1.1 Annual Publication Count

Total publication counts for DL and ML studies are 1407 and 2245 respectively. Evolution of these numbers is given in Fig. 1a in yearly basis. We started the plot from year 2014 since previous publication counts are less than 10. Popularity of ML in 5G studies starts from this year with a steep increase, while DL counterpart receives scientific interest with a delay of ~3 years. However, rate of DL studies to ML studies given in the right panel is in an increasing trend, still being below ML counterpart. Tendency of this graphic says that ML-related studies will remain more popular for some coming years, despite the increasing interest for DL.

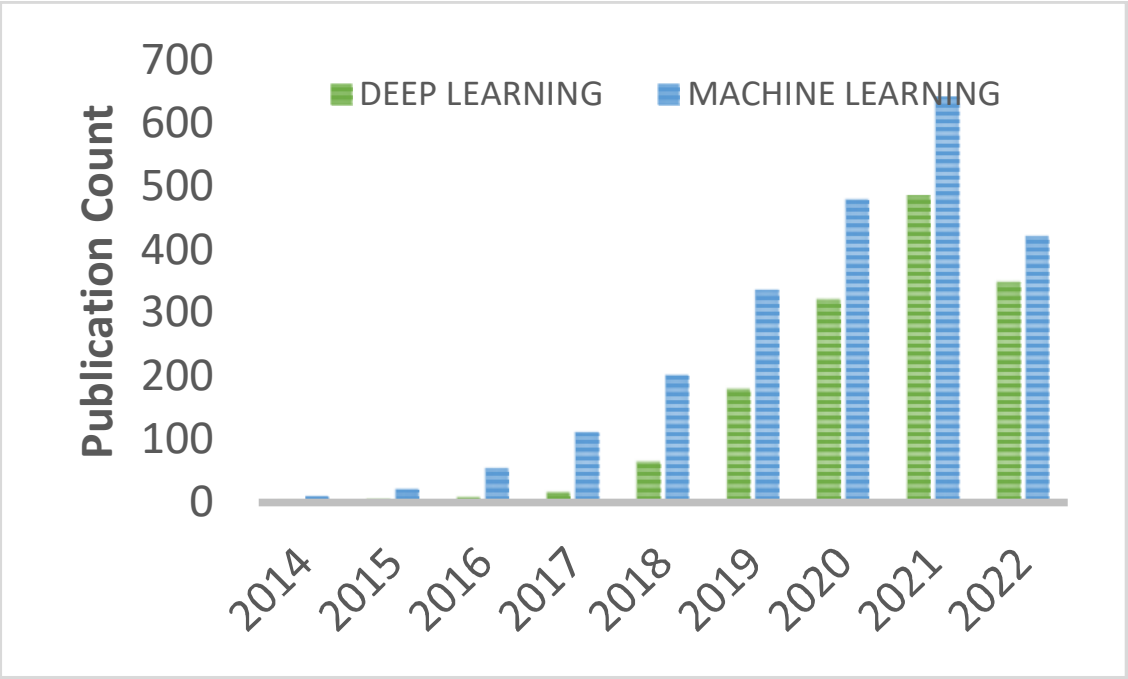

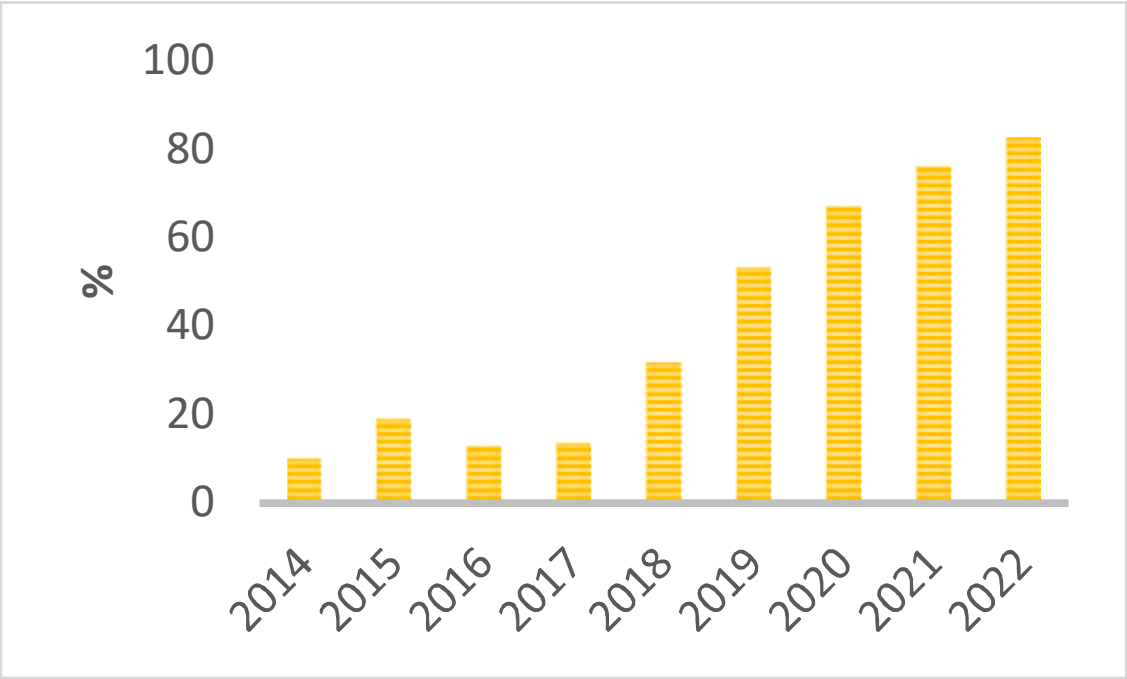

**Fig. 1. a) Evolution of publication count for years. b) DL/ML publication rate in percent.**

#### 3.1.2 Document Types

Researchers prefer publishing in different document types such as articles, conference proceedings, books etc. Dispersion of the 5G-related publications for document types is given in Fig. 2.

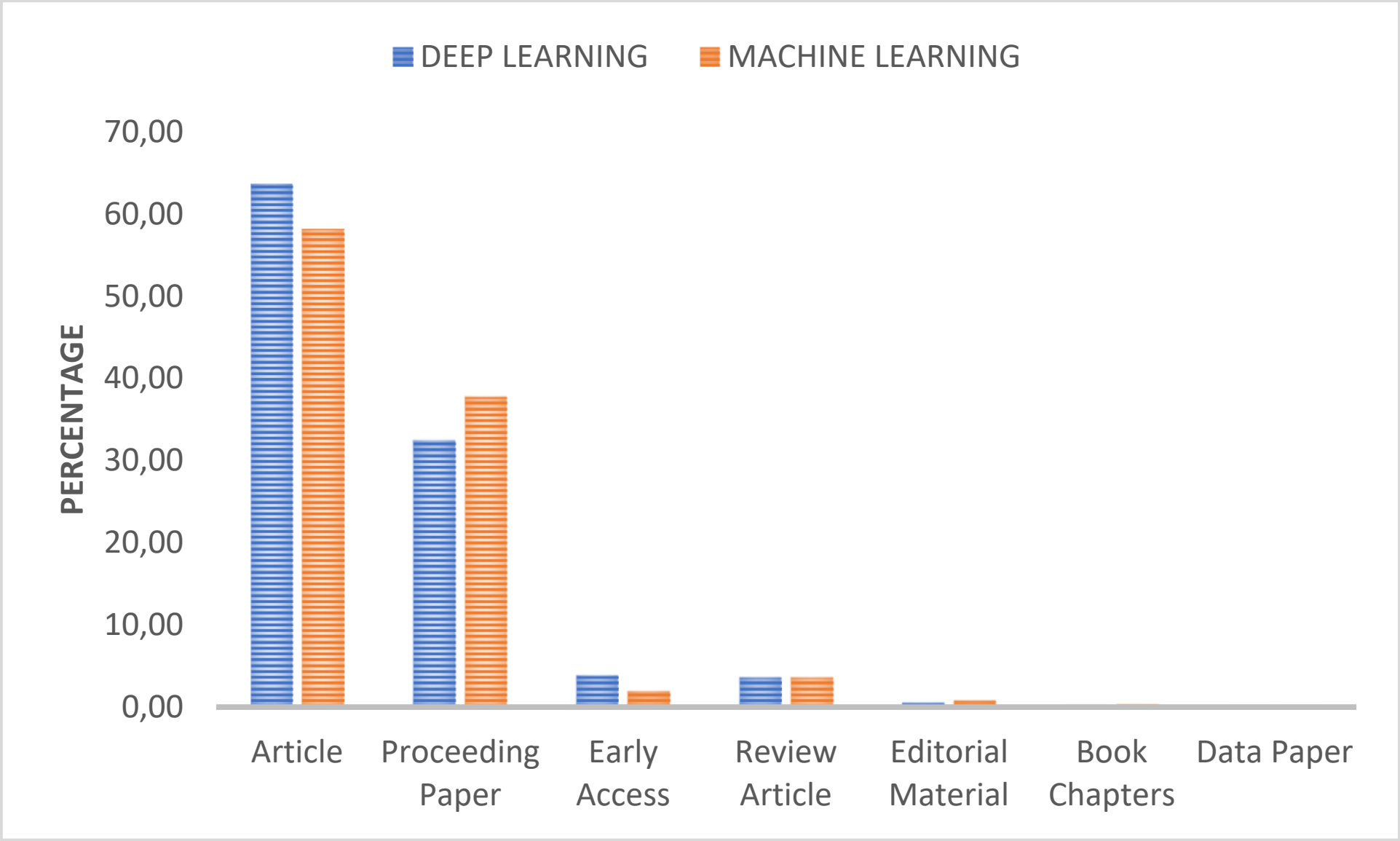

**Fig. 2. Distribution of DL and ML studies in 5G scope according to document types.**

Fig. 2 outlines that DL studies have slightly more preference to be published as articles, while ML studies have greater percentage for conference proceedings. Both streams have very low frequency to publish review papers (3.6%) with very close rates. Last 3 options have very low frequencies (<1%) for meaningful inferences.

### 3.1.3. Publication Titles (Sources)

Authors tend to publish papers with the most suitable source for their studies, also considering the acceptability in a reasonable period. Therefore, publication titles is an important consideration for researchers in any field. Fig. 3 presents publication rates for sources, while the second panel of this figure presents the preferential acceptance rate for DL studies compared to ML counterpart. This preference of publishing DL studies in comparison with ML studies in quantized by calculating a $t_{DL}$ (*tendency of publishing DL)* percentage metric calculated as below:

$$t_{DL} = \frac{100(p_{DL} - p_{ML})}{p_{ML}} \tag{1}$$

Where of publishing DL and ML studies for the given source. This metric is later applied to other attributes in coming subsections. Please note that $p_{DL}$ and $p_{ML}$ are calculated internally (i.e. $p_{DL}(IEEE\ Access)$ is the probability of publishing with IEEE Access journal for only DL studies, which can be calculated by dividing the number of papers published in this journal by the total number of *DL* studies).

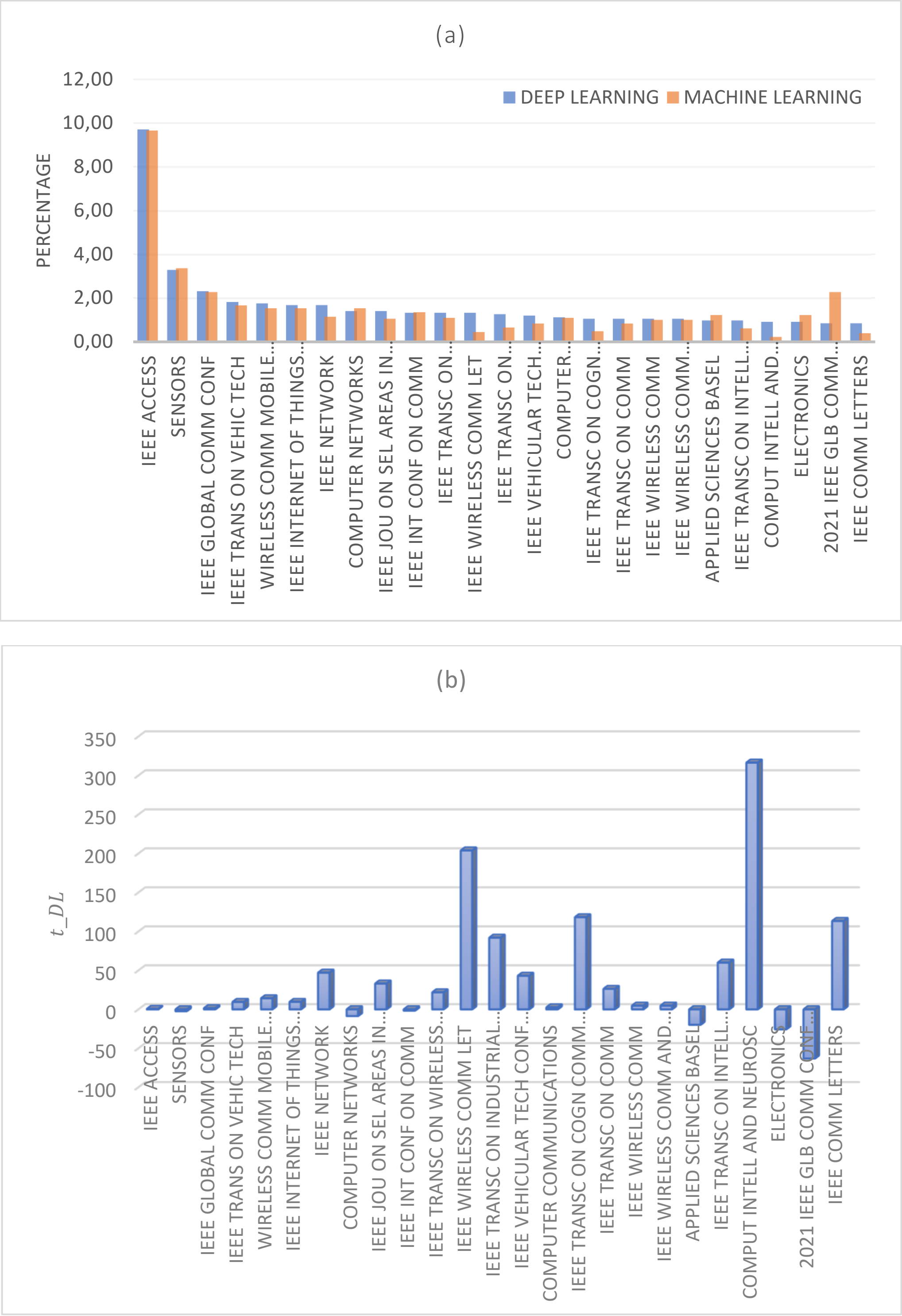

**Fig. 3. a) Publishing rates of DL and ML studies according to sources. b) Tendency of sources to publish DL studies compared to ML studies.**

Fig. 3 indicates that IEEE Access (~9.7%) and Sensors (~3.3%) are the two journals to most frequently publish 5G-related studies, while the next source with greatest publication rate is IEEE Global Communications Conference. 18 of the top 25 sources are IEEE journals or conferences. Panel (b) indicates that Computational Intelligence and Neuroscience, IEEE Wireless Communications Letters, IEEE Transactions on Cognitive Communications and Networking, IEEE Communications Letters and IEEE Transactions on Industrial Informatics are the most selective journals (>50%) for DL-based studies with respect to ML-based studies in 5G networks domain. $t_{DL}$ metric indicating this tendency is in $\pm 2\%$ range for the first 3 sources with highest volumes, exhibiting no significant preference for DL or ML-based studies.

### 3.1.4. Research Areas

Though primarily belonging to telecommunication scope, research area attribute for DL or ML-driven 5G studies exhibits a broad range of scopes. As presented in Fig. 4, there exists a dominance of 3 areas as Engineering, Telecommunication and Computer Science. Since sum of internal percentages provided by WoS exceed 100%, papers emerge to be multi-labeled as belonging to more than one research areas.

As both panels indicate, Engineering scope has a very close rate for both DL and ML approaches, while telecommunication scope exhibits limited negative (-5.71%) and computer science scope exhibits limited positive (1.39%) selectivity for DL approach. Lower panel dedicated to tendency of publishing DL-based studies show greater fluctuations (>50%) for low-volume research areas. Among these research areas, most positive tendencies for DL approach are acoustics, neurosciences and neurology, mathematical computational biology and remote sensing. No significant negative selectivity for DL approach is observed.

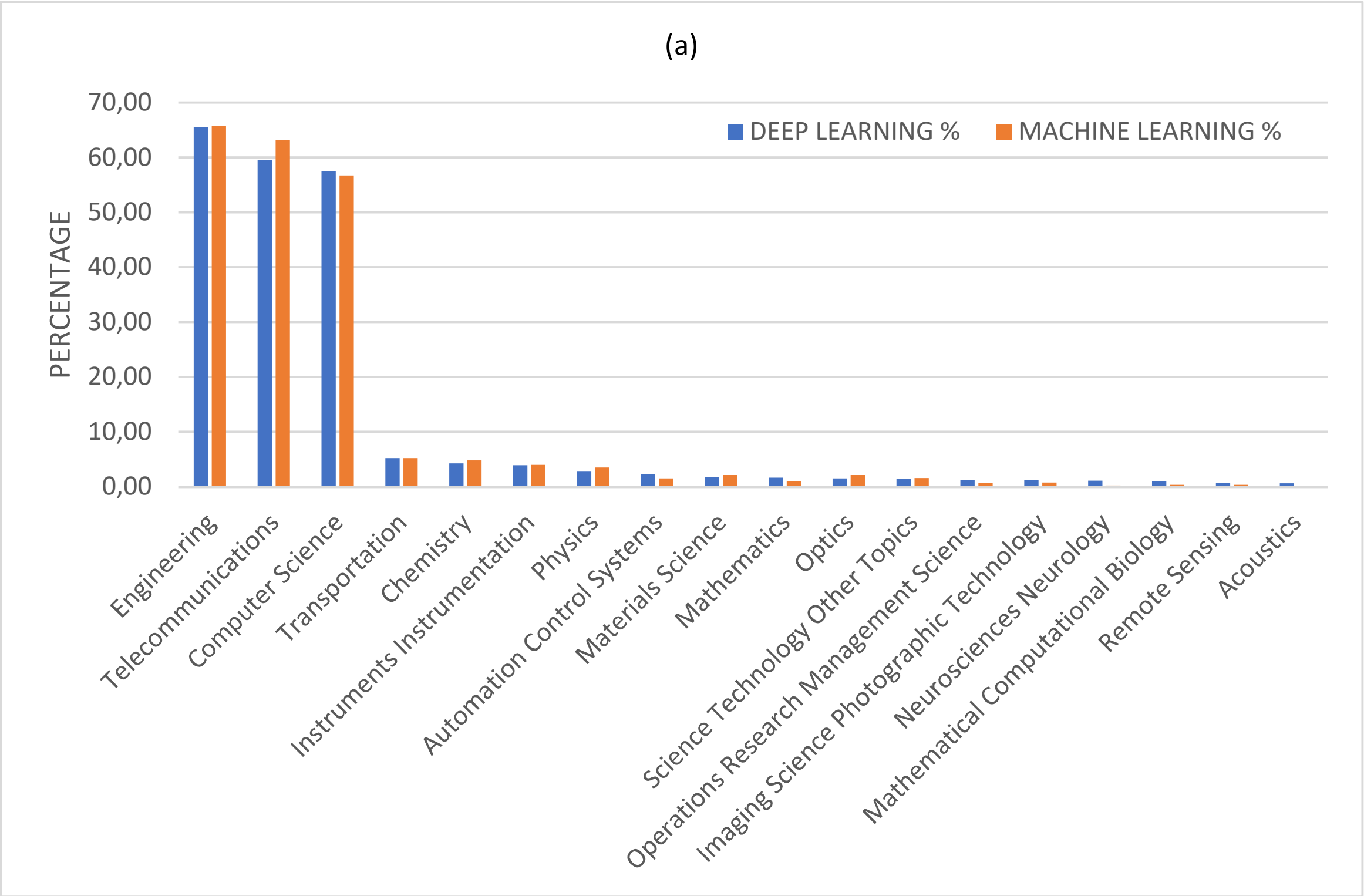

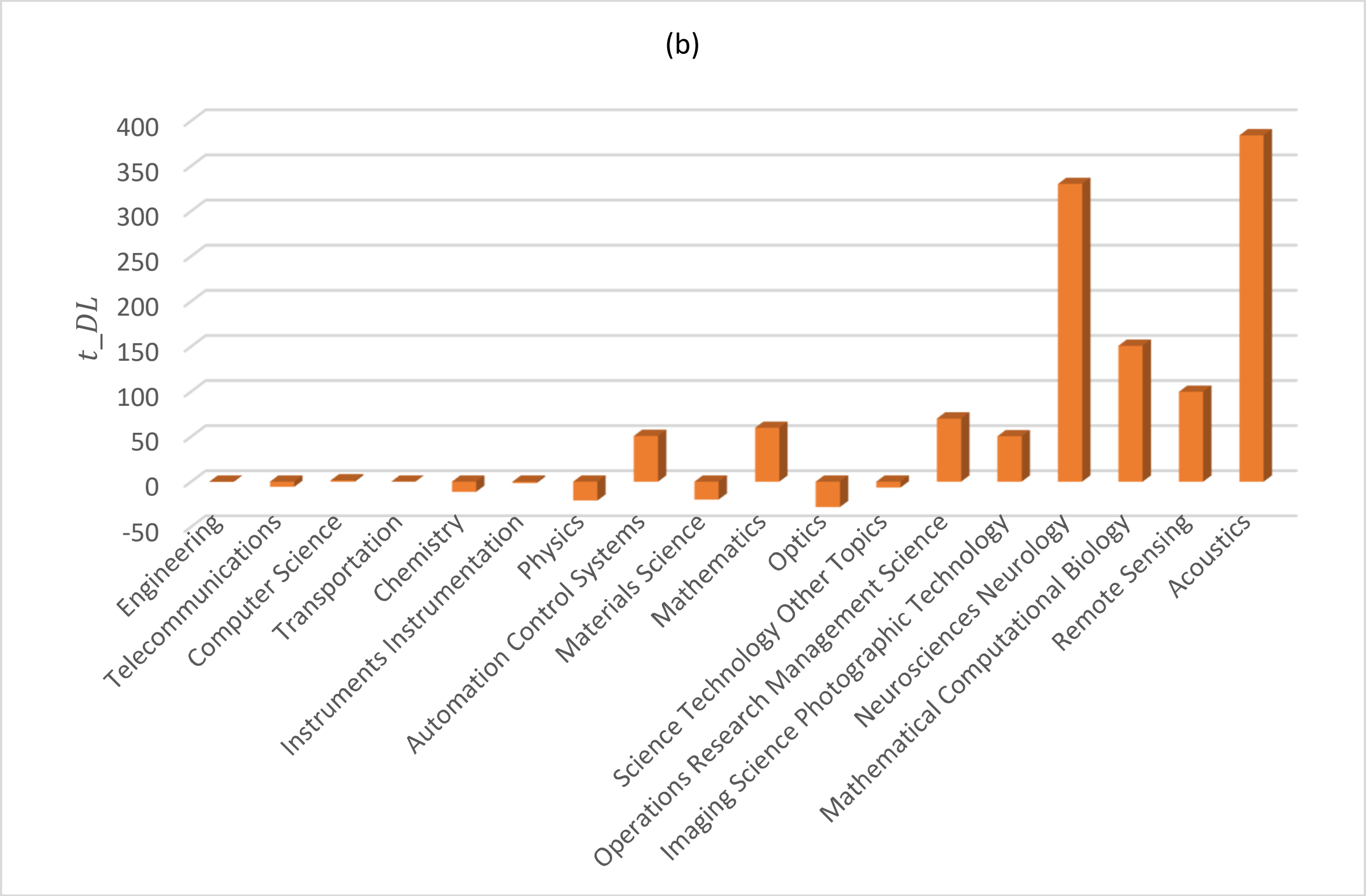

**Fig. 4. a) Publishing rates of DL and ML studies according to research areas. b) Tendency of research areas to favor DL studies with respect to ML studies.**

### 3.1.5. Countries / Regions

A noteworthy outcome of our comparative approach for DL and ML approaches in 5G-related studies is that regional tendencies to favor DL or ML-based approaches are apparent. As given in Fig. 5, Republic of China has the highest volume to publish in this domain, at least 2 times far from its nearest follower (USA). 30% of positive selectivity for DL-based approaches is apparent for China, while this positive tendency for DL is also observed with smaller rates for the next 4 countries with highest publication volumes.

Countries with highest selectivity for DL approaches are Taiwan, Egypt, Qatar, Vietnam, China, Saudi Arabia, Norway, England, Russia, with tendency rates ranging from 63 to 21.5%. Contrarily, countries with highest negative selectivity for DL approaches (meaning they favor ML instead) are Greece, Sweden, Arab Emirates, Malaysia, Germany, France, Spain, Italy, Finland, Japan, with tendency rates ranging from -67% to -20%. Panel c of the same figure also presents dispersion of DL-related 5G studies worldwide, in a map chart visualization.

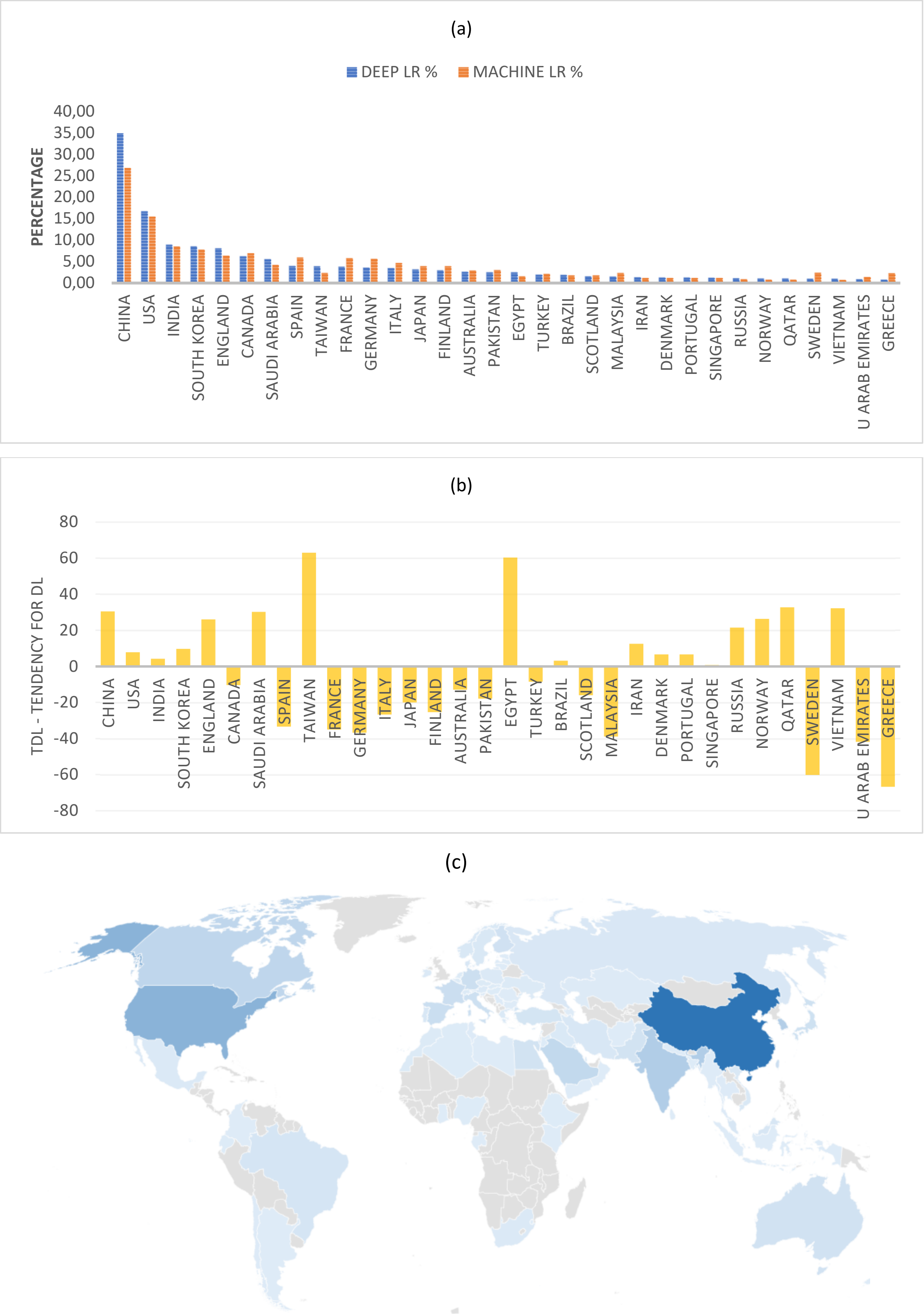

**Fig. 5. a) Publishing rates of DL and ML studies according to countries/regions. b) Tendency of countries/regions to favor DL studies with respect to ML studies. c) Map chart demonstrating interest in publishing DL-based 5G networks papers.**

### 3.1.6. WoS Index

One of the most important considerations in scientific publishing is the indexing services for a given paper or source. Dispersion of indexing services of 5G studies regarding their DL or ML-based methodology is given in Fig. 6.

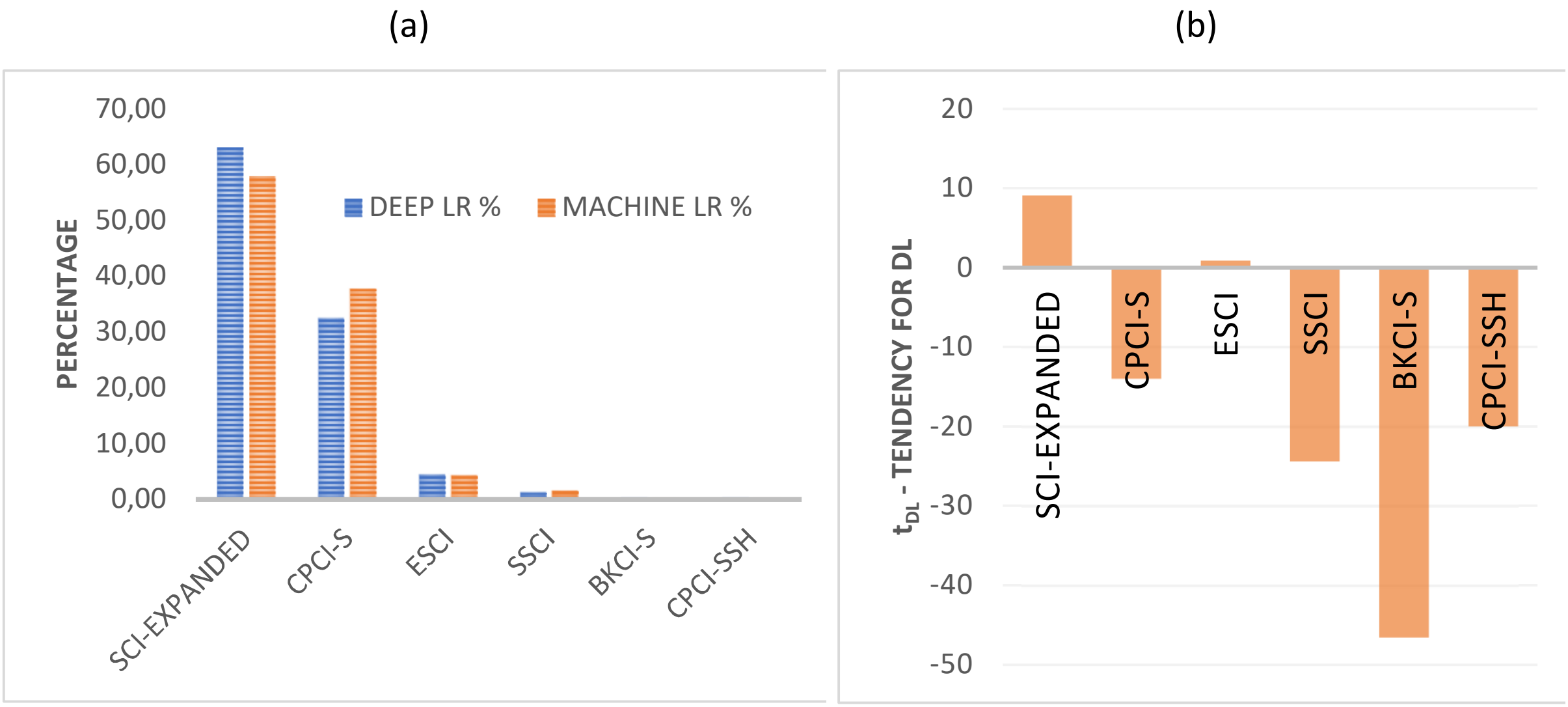

**Fig. 6. a) Publishing rates of DL and ML studies according to WoS indexes. b) Tendency of WoS indexes to favor DL studies with respect to ML studies.**

Fig. 6 indicates that DL has a greater tendency to be indexed by SCI-Expanded, that is also observed in a minor rate for ESCI index. Other indexes dominantly favor publishing ML approach in 5G rather than DL. This may be an indicator that SCI-E and ESCI indexed journals tend to seek for DL approaches in 5G network studies in their editorial progress.

### 3.2. Evaluating Scientific Impact

The current section focuses on scientific impact measured with citation and usage counts for the published papers. The availability of fields in WoS dataset related with citation and usage count along with the previously mentioned attributes enables extraction of impact-based inferences.

Since citation count is a yearly increasing attribute, it is normalized as dividing by the age of the publication in the following results. Therefore, "normalized citation count" is considered as the metric to measure the scientific impact, labelled as "yearly normalized citations". This attribute is also aggregated with respect to years, leading a yearly plot of evolution of citation metrics.

### 3.2.1. Average Normalized Citation Count vs. Years

As given in panels *a* and *b* of Fig. 7, DL studies (2.256) have received slightly more average citations compared to ML studies (2.118) in 5G. While this difference in citation performance remains as 6.5%, a greater difference in usage counts 15.8% emerges, as an indicator that reading-to-citing rate is lower for DL studies compared to ML studies.

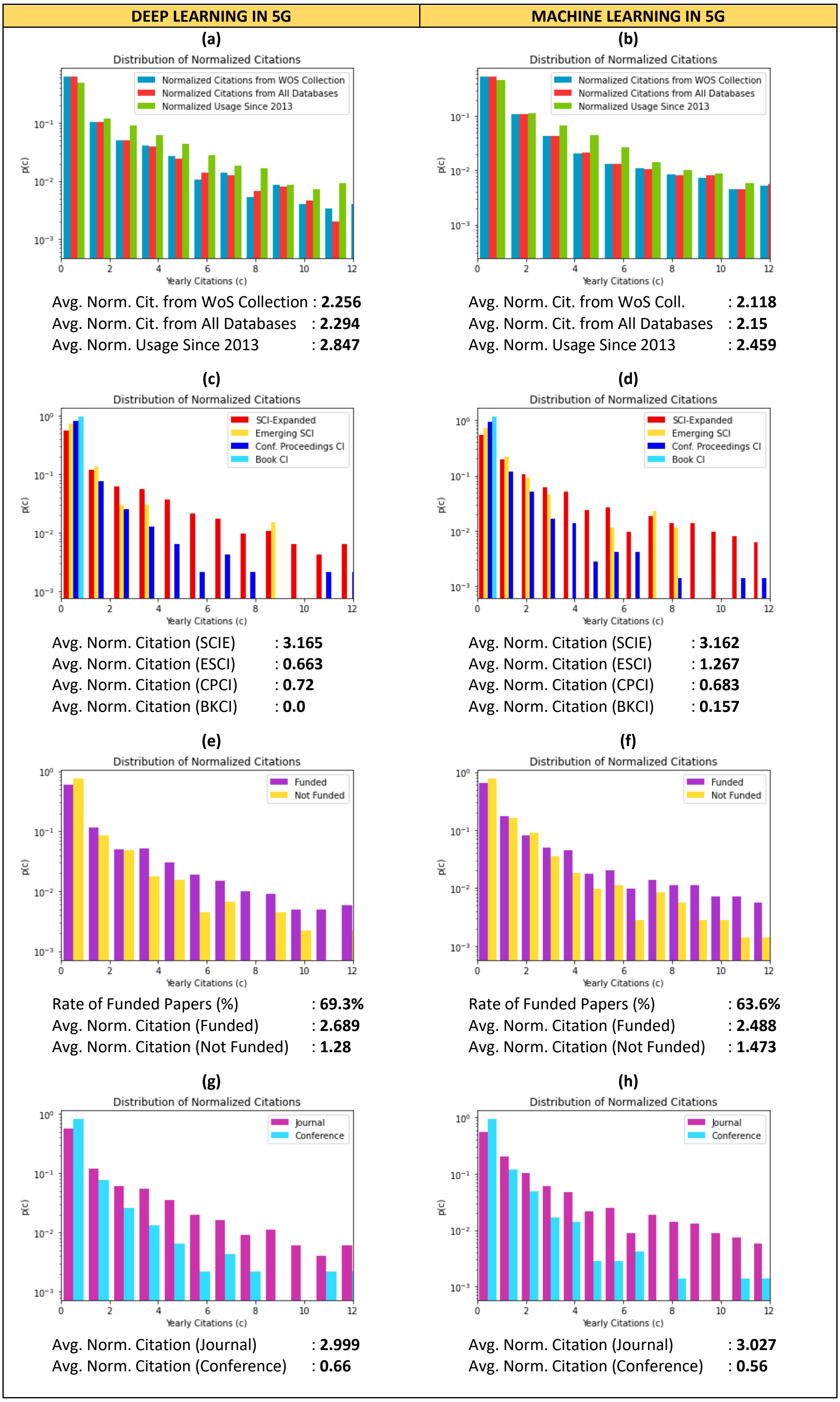

**Fig. 7. Distributions for (a,b) normalized citation and usage counts, (c,d) WoS Index, (e,f) Funding availability, (g,h) Journal or conference publication. For all groups, left panels represent deep learning and right panels represent machine learning approaches. Average values are also provided below the panels.**

This low convertibility from reading to citation is also observable in the distribution plots, since green bars in DL side are noticeably greater than red and blues. As a general conclusion, one can say that DL approach has slightly greater impact compared to ML in 5G domain.

### 3.2.2. Average Normalized Citation Count vs. Indexing Information

Panels *c* and *d* are dedicated for this title, outlining very similar citation patterns and counts for SCI-Expanded indexed papers of DL and ML-based 5G studies. However, ESCI indexed ML-related papers and CPCI indexed DL-related papers have noticeably greater citation performances compared to their peers. This is an indicator that interest in DL side is more conference-driven and ML side more journal paper driven.

### 3.2.3. Average Normalized Citation Count vs. Funding Availability

DL side consists of 69.3% funded papers while this rate is 63.6% in ML side, as traced in panels *e* and *f* of Fig. 7. This indicates approximately 10% higher funding availability for DL side. Funded DL studies receive more citations against funded ML studies, whereas unfunded ML studies receive more citations against unfunded DL studies. For both approaches, funding availability boosts citation performance approximately 2 times.

### 3.2.4. Average Normalized Citation Count vs. Publication Type

Panels *g* and *h* of Fig. 7 outline that, citation performance of DL and ML-based 5G *journal papers* are very close (1% greater for ML), while DL-based *conference papers* have a potential to receive 18% greater citations. This outcome also supports our previous inference that DL side receives more attention through conference events. For both streams, journal papers have 5-6 times greater potential to receive citations compared to conference papers. This may be a key reason why researchers should aim publishing with journals.

### 3.2.5. Timely Histogram for Normalized Citations

Another important consideration beyond average normalized citations is the dispersion of these citations over years. Patterns in yearly distributions may give more information about timeliness of the papers' popularity. Fig. 8 presents histograms for averaged normalized citations over years, for both DL and ML streams. These distributions indicate that citation pattern of ML-based studies is dominated by some pioneering studies from year 2014, following a uniform-like histogram for the following years. Normalized citation histogram DL-based studies differ from the ML side, with a bell-like structure with peak values from years 2017-2019.

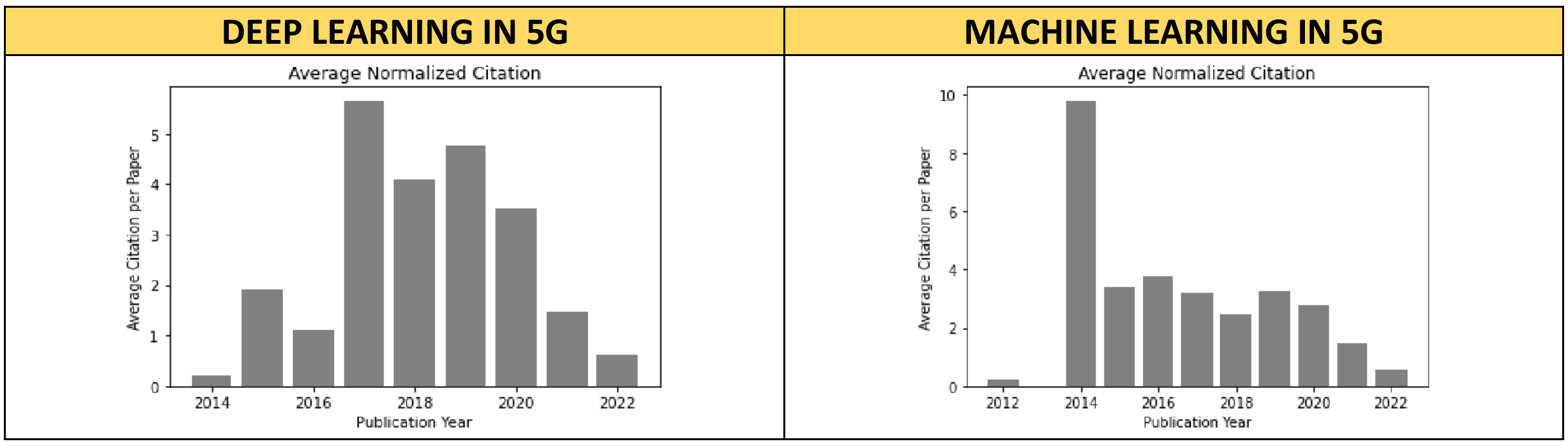

**Fig. 8. Distributions for average normalized citation for years. Panels represent DL and ML approaches in 5G respectively.**

Since 5-year impact is an important measure in bibliometrics, last 5 bars of both distributions may be a good indicator for the future of the two streams. Numerically aggregating last 5 years, we get average citation counts as 2.89 for DL and 2.11 for ML-based studies. Therefore, we can conclude that DL-induced studies for 5G networks will be advantageous by the means of scientific impact in the future.

## 4. CONCLUSIONS

As a result of a bibliometric view on DL and ML-induced studies on 5G Networks from WoS database, following conclusions can be listed.

- AI-related 5G Network studies are still mostly conducted with ML approaches more than DL approaches. However, employment of DL is in an increasing trend, having reached to ~83% of ML studies in 2022.
- DL studies are published as articles in a higher rate, while ML studies have higher rate to be published as conference proceedings. However, DL studies published in conferences and ML studies published in journals contrarily tend to receive greater citations.
- IEEE Access and Sensors journals hold for 13% of all AI-related 5G studies with no significant selectivity for DL or ML-driven papers. 18 of the top 25 sources are IEEE journals or conferences in this scope.
- There exists more selectivity for DL rather than ML, which is observed in a variety of publication sources such as Computational Intelligence and Neuroscience, IEEE Wireless Communications Letters, IEEE Transactions on Cognitive Communications and Networking, IEEE Communications Letters and IEEE Transactions on Industrial Informatics.
- Telecommunication scope exhibits limited negative (-5.71%) and computer science scope exhibits minor positive (1.39%) selectivity for DL approaches in 5G. Significant publishing tendencies (>50%) for DL approach are acoustics, neurosciences and neurology, mathematical computational biology, robotics and remote sensing.
- Rep. of China has the highest volume to publish in this domain, at least 2 times far from its nearest follower (USA).

- Countries with highest selectivity for DL approaches are Taiwan, Egypt, Qatar, Vietnam, China, Saudi Arabia, Norway, England, and Russia. Contrarily, countries with highest selectivity for ML approaches are Greece, Sweden, Arab Emirates, Malaysia, Germany, France, Spain, Italy, Finland, and Japan.
- SCI-E indexed journals tend to seek for DL approaches in 5G network studies with a noticeable rate, while BKCI, SSCI and CPCI indexed journals favor ML approaches instead.
- DL studies in 5G domain receive 15.8% higher reading and 6.5% higher citation performance compared to ML studies.
- DL-induced papers receive funding in 10% higher rate compared to ML part, while funded papers in both sides receive 2 times more citation.
- Journal papers dedicated to DL and ML approaches in 5G exhibit very similar citation performance, while DL-based conference papers have a potential to receive 18% greater citations.
- Having focused to last 5 years, average citation counts emerge as 2.89 for DL and 2.11 for ML-based studies, indicating that DL side exhibits greater 5-year impact. This may be an indicator that interest to DL-based approaches will be greater in coming years.